\documentclass{iau_jrad}

\usepackage{graphicx}
\usepackage[suffix=]{epstopdf}
\usepackage{natbib}

\usepackage{aas_macros}

\title[Davenport: Shape of Flares] 
{The Shape of M Dwarf Flares in \\Kepler Light Curves}

\author[Davenport]   
{James R. A. Davenport$^{1,2}$}

\affiliation{
$^1$NSF Astronomy and Astrophysics Postdoctoral Fellow\\
Department of Physics \& Astronomy, \\
Western Washington University, Bellingham, WA 98225 \\ 
email: {\tt James.Davenport@wwu.edu}\\
[\affilskip] 
 $^2$Department of Astronomy, \\
 University of Washington, Box 351580, Seattle, WA 98195
 }

\pubyear{2015}
\volume{320}  
\jname{Solar and Stellar Flares and their Effects on Planets}
\editors{A. G. Kosovichev, S. L. Hawley, P. Heinzel}
\begin{document}

\maketitle

\begin{abstract}
Ultra-precise light curves from Kepler provide the best opportunity to determine rates and statistical properties of stellar flares. From 11 months of data on the active M4 dwarf, GJ 1243, we have built the largest catalog of flares for a single star: over 6100 events. Combining 885 of our most pristine flares, we generated an empirical white-light flare template. This high-fidelity template shows a rapid initial rise, and two distinct exponential cooling phases. This template is useful in constraining flare energies and for improved flare detection in many surveys. Complex, multi-peaked events are more common for higher energy flares in this sample. Using our flare template we characterize the structure of complex events. In this contributed talk, I presented results from our boutique study of GJ 1243, and described an expanded investigation of the structure of complex flares and their connection to solar events.
\end{abstract}

\firstsection 

\section{Introduction}
Photometric data from the Kepler mission \citep{borucki2010} has ushered in a new era for time domain studies of stellar magnetic activity, with starspots and flares being observed for many active stars \citep[e.g.][]{walkowicz2011}. These long duration light curves for thousands of solar-like stars enable the first complete statistics of rare ``superflare'' events \citep{shibayama2013}, and of flares from stars across a wide range of masses \citep{balona2015a}.

We have recently published a detailed study on the rates and properties of flares for a handful of active and inactive Kepler M dwarfs \citep{hawley2014}. Flares in this work were manually identified in the Kepler short-cadence (1-minute) light curves, and have revealed multiple flare events on stars with no detectable starspot modulations in their light curves or chromospheric H$\alpha$ emission in their spectra. Follow-up work on the active M5+M5 binary system GJ 1245AB has also revealed unique starspot properties and flare rates for both stellar components \citep{lurie2015}. 

\begin{figure}
\begin{minipage}[c]{0.6\textwidth}
\includegraphics[width=3in]{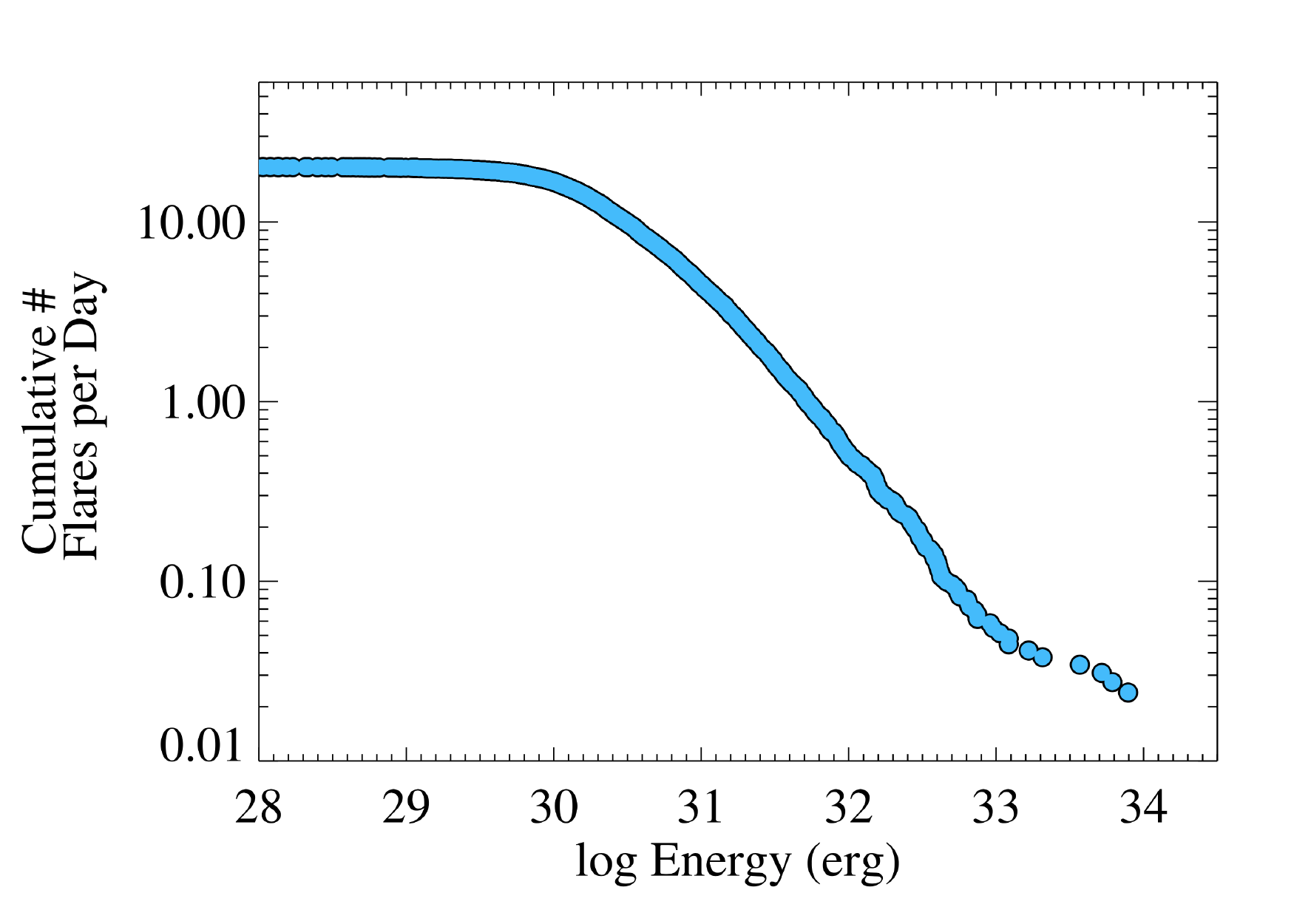} 
\end{minipage}
\begin{minipage}[c]{0.35\textwidth}
\caption{Cumulative flare frequency distribution for the 6107 flares observed on GJ 1243, sorted from largest to smallest energy events. A single power-law slope is seen between log E $\sim30.5$--32.5 erg. The turnover at low energy is at least partly due to decreasing detection efficiency for low amplitude flares, and possibly a real change in power-law \citep[see][]{hawley2014}. The break from a single power-law at the high end is due to sample incompleteness for these rare events.}
\label{fig:ffd}
\end{minipage}
\end{figure}

The sample from \citet{hawley2014} also included the rapidly rotating M4 dwarf, GJ 1243.  This highly active star was the focus of both a detailed starspot tracking analysis that revealed slow differential rotation using 4 years of long-cadence data \citep{davenport2015b}, and a study of the temporal morphology of flares from 11 months of short-cadence data \citep[][hereafter D14]{davenport2014b}. 
The construction of the flare sample is described fully in D14. Briefly, users manually inspected each of the 11 months of Kepler short-cadence data, picking the start and stop time of candidate flare events using {\tt FBEYE}\footnote{Available online at http://github.com/jradavenport/FBEYE}. Flares were then selected as candidates from overlapping regions of time that multiple users had selected. This produced the largest sample of flares ever amassed for a single star (besides the Sun), with a total of 6107 flare events in 11 months of Kepler data. 
The ensemble cumulative flare frequency distribution for this sample is shown in Figure \ref{fig:ffd}. 
Here I describe additional aspects of flare event morphology from GJ 1243, as well as future avenues for research with this unique sample.

\section{The Flare Template}

\begin{figure}[!h]
\begin{minipage}[c]{0.6\textwidth}
\includegraphics[width=3in]{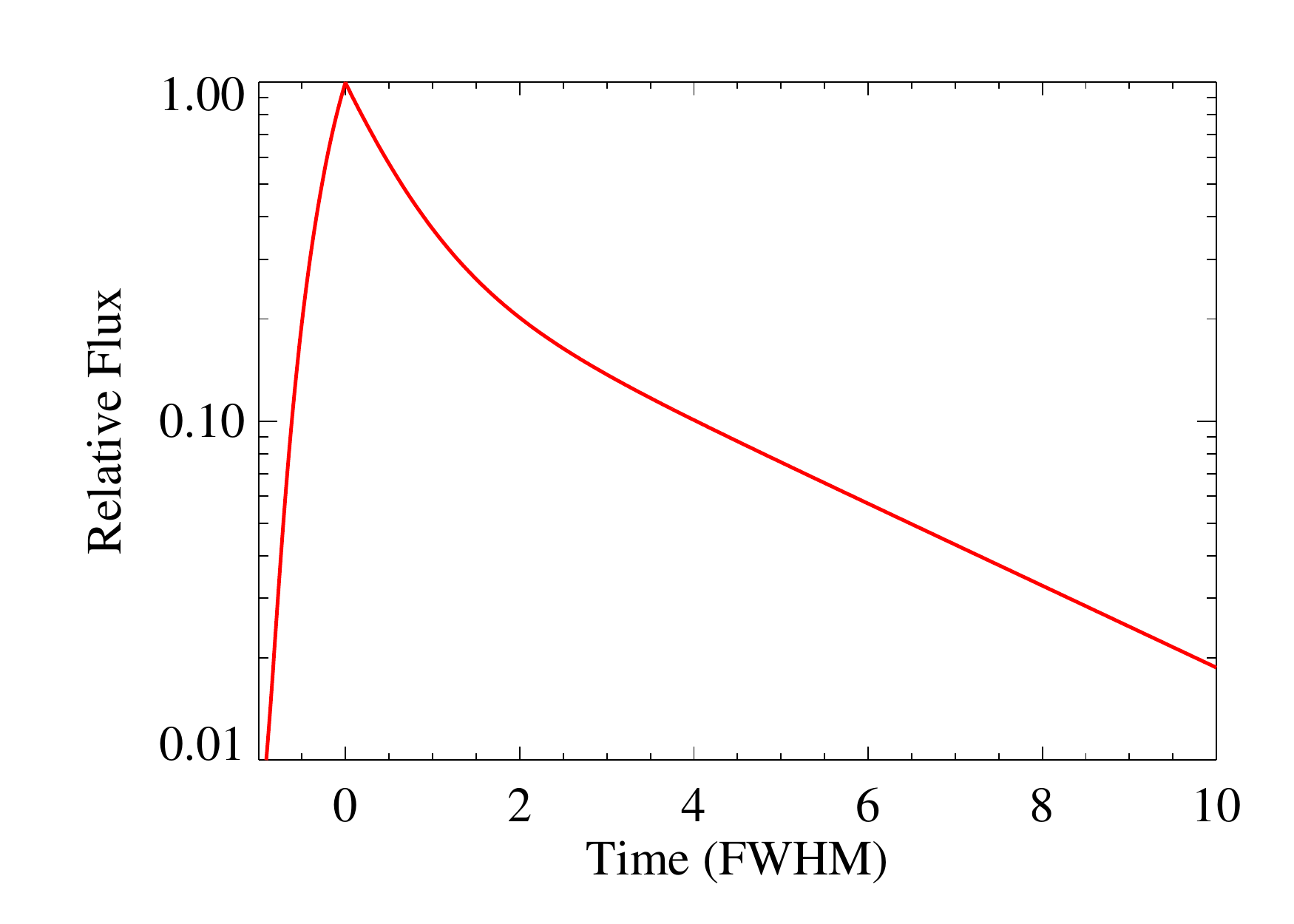}
\end{minipage}
\begin{minipage}[c]{0.35\textwidth}
\caption{The analytic flare model from \cite{davenport2014b}, shown using logarithmic flux scaling. Three regimes of the flare are visible: the rapid rise phase, and two exponential decay phases.}
\label{fig:model}
\end{minipage}
\end{figure}

As described in D14 and \citet{hawley2014}, the sample of flares from GJ 1243 was used to produce an empirical flare template. This template was generated from only flares that users had selected as having a ``classical'', single-peaked morphology, and with durations or 10 minutes of greater. A sample of 885 flare events passed these criteria, and were normalized by their amplitudes and characteristic timescales (defined as the flare full-width at half-maximum, or FWHM). These normalized flares were then median combined to produce the empirical flare shape. The rise and decay phases of the median-combined flare were then fit by D14 with analytic functions, producing the analytic flare model shown in Figure \ref{fig:model}. 

\section{Complex Flares}

The flare template described above can be used to objectively identify and decompose complex flare events.  D14 demonstrated this procedure using an iterative approach to deconstruct complex flares, re-fitting each event with additional flare components. With each additional flare component that was included, the Bayesian Information Criteria was used to evaluate the quality of the fit while penalizing the increased degrees of freedom in the model. Code that demonstrates this iterative approach is available online\footnote{http://github.com/jradavenport/flare-fit}, but users are encouraged to build their own fitting routines around the empirical flare template.

\begin{figure}[]
\begin{center}
\includegraphics[width=2.5in]{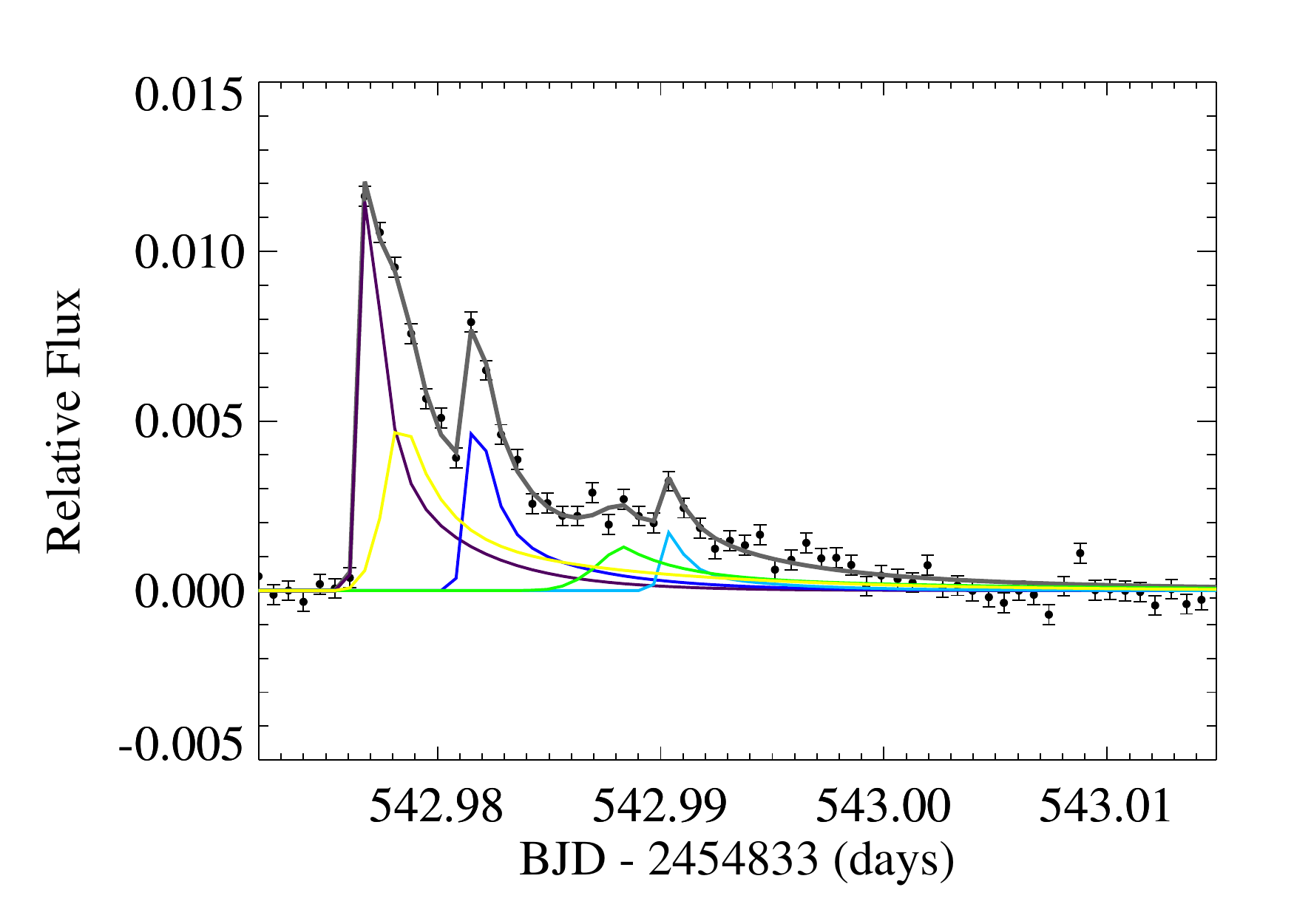}
\includegraphics[width=2.5in]{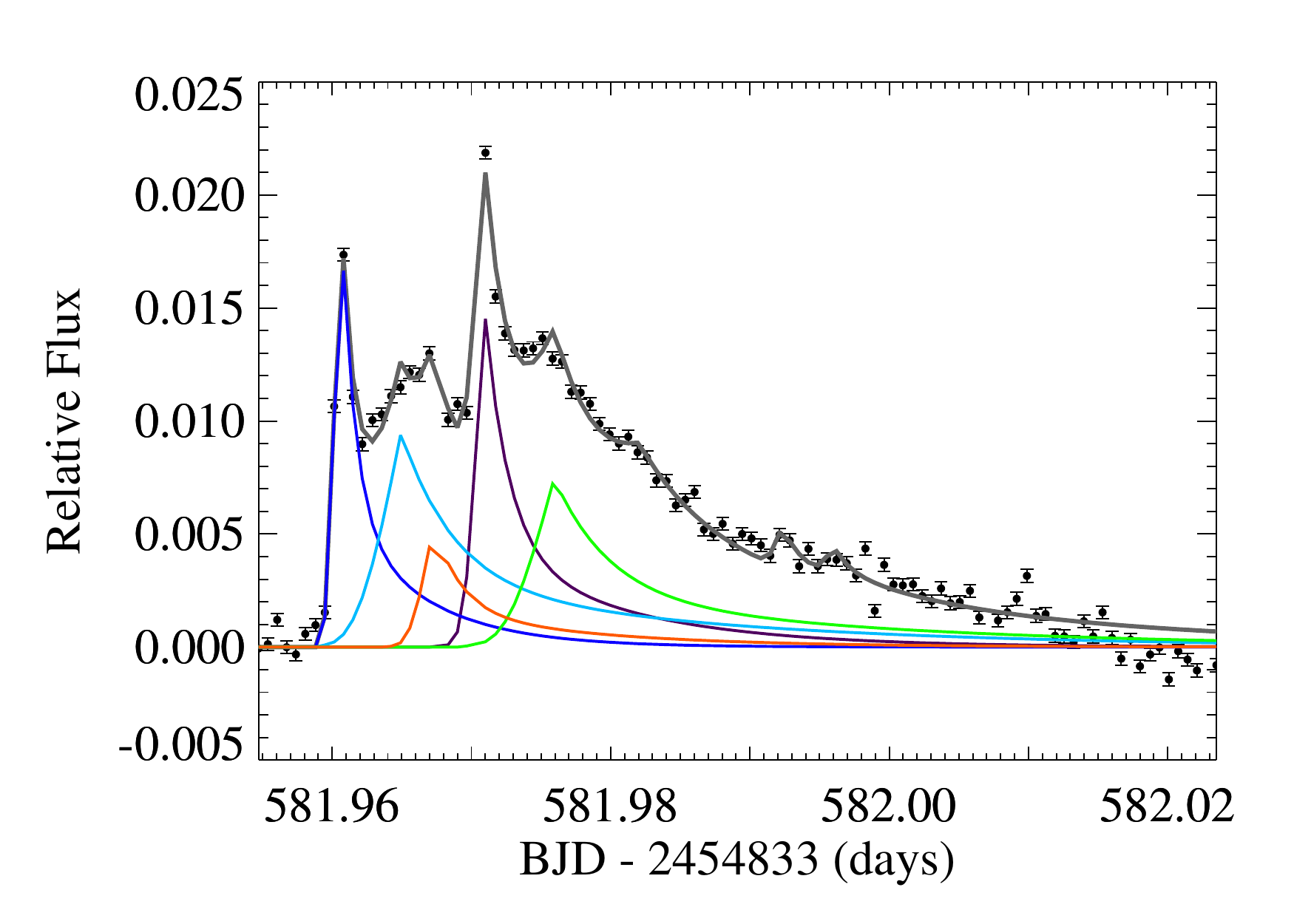}
\caption{Two examples of complex flares deconstructed using the iterative fitting approach from D14. A relatively simple complex flare (left) is modeled well by 5 flare components. A highly complex event (right) is correctly identified as a complex flare by the D14 algorithm, but the preferred 5 component fit does not precisely reproduce every feature in the event.}
\label{fig:cplx}
\end{center}
\end{figure}

Two examples of this iterative fitting are given in Figure \ref{fig:cplx} for complex flares from GJ 1243. These events show examples of the fitting procedure successfully identifying and decomposing complex flares. The first event shows a relatively simple complex event, with one primary peak and several distinct secondary flares with smaller energies that are well matched using the template. The second example in Figure \ref{fig:cplx} shows a highly complex event with multiple major peaks, and a very gradual decay that may show evidence of small amplitude oscillations as seen in other complex solar and stellar flares \citep[e.g.][]{inglis2015}. The iterative fitting routine from D14 correctly identifies this event as complex, preferring a model with 5 components. The uniqueness of the fit in this case is questionable, however, and the iterative fitting fails to reproduce the small oscillations in the flare decay.

\begin{figure}
\begin{minipage}[c]{0.6\textwidth}
\includegraphics[width=3in]{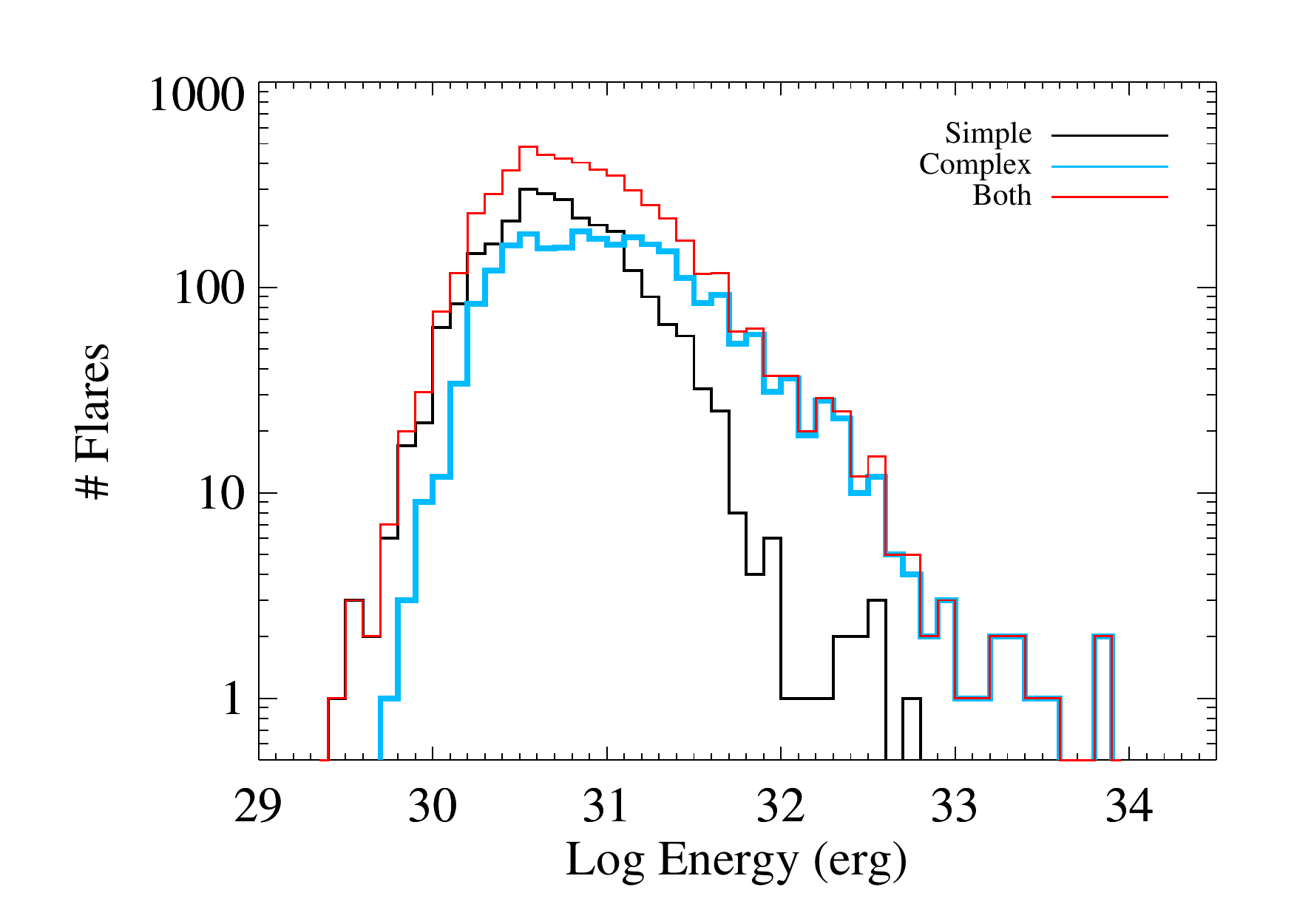} 
\end{minipage}
\begin{minipage}[c]{0.35\textwidth}
\caption{Distribution of flare energies for both complex (blue) and classical (black) events. Complex events dominate the flare census at high energies. Note that while complex flares are seen at all event energies, they dominate higher energy events.}
\label{fig:hist}
\end{minipage}
\end{figure}

As D14 note, the fraction of complex events dominates the flare census at high energies. In Figure \ref{fig:hist}, I present the distribution of flare event energies for both the complex and classical flares. \citet{balona2015a} found 32\% of flares across many spectral types exhibited ``complex'' behavior. From our total sample of 6107 flares, human classification indicated only 15.5\% of events were complex. However, for events with durations of 10 minutes or longer (3737 flares), fitting using the iterative approach from D14 produces a total complex flare rate of 30.5\% (1141 flares).

\begin{figure}
\begin{minipage}[c]{0.6\textwidth}
\includegraphics[width=3in]{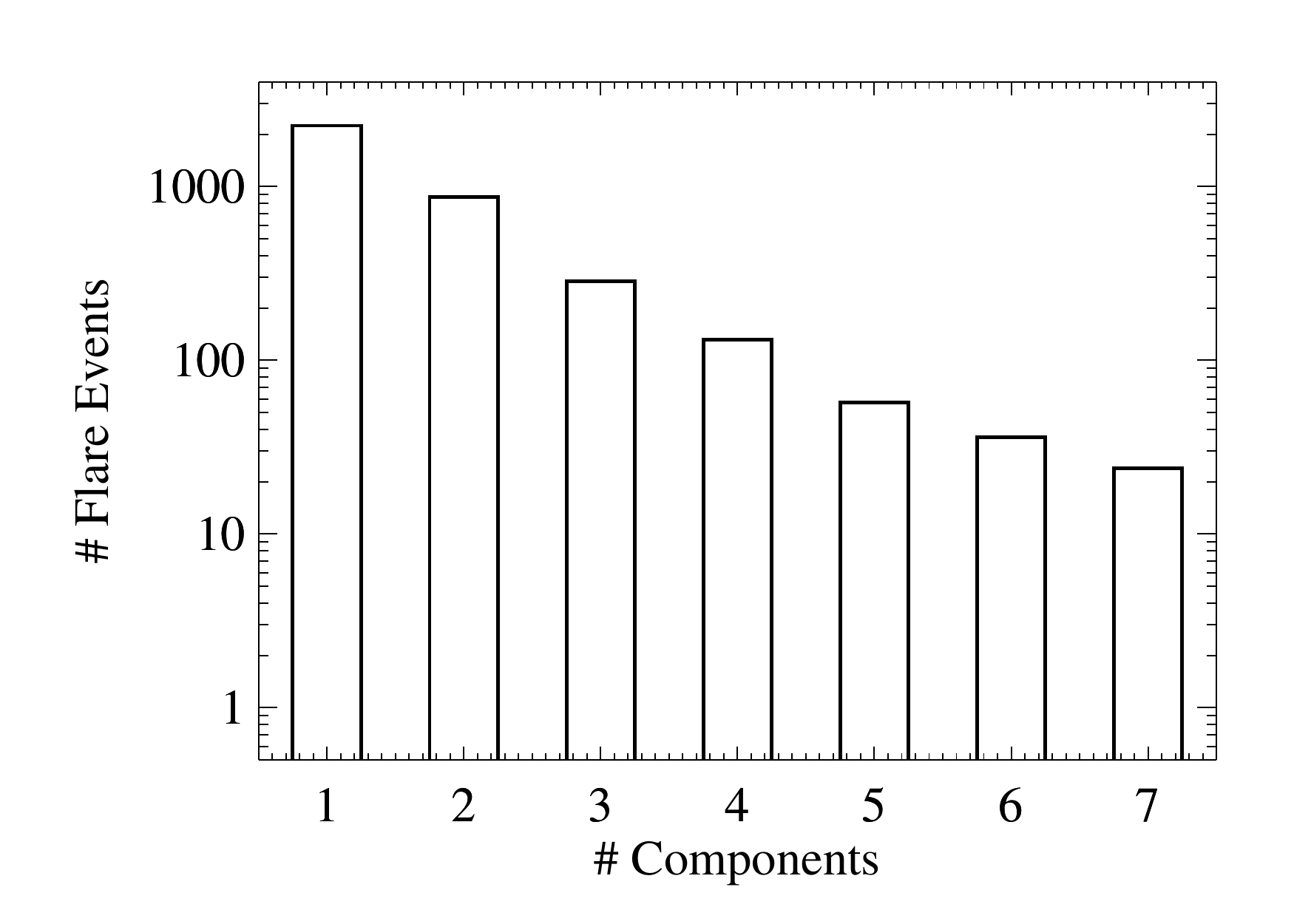} 
\end{minipage}
\begin{minipage}[c]{0.35\textwidth}
\caption{Distribution of the number of component flares fit in the 3737 flare events with durations of 10 minutes or longer. In total, 1141 events (30.5\%) were classified as ``complex'' having more than 1 component flare.}
\label{fig:ncomp}
\end{minipage}
\end{figure}

\section{Degree of Complexity in Complex Flares}

From the sample of 1141 complex flares identified by D14, we may begin to construct a statistical picture of the rates, compositions, and ``degree of complexity'' for these events.
In Figure \ref{fig:ncomp}, I present the distribution of the number of components (analytic flare models) fit to these complex flare events, which is the most basic measure of the degree of event complexity. An exponentially decreasing number of components is found in this distribution, which I postulate is related to the distribution of sizes and geometries of the active regions as they flare and trigger secondary flare events. This may also be related to ``sympathetic flaring'' observed on the Sun \citep[e.g.][]{schrijver2015}.
Comparable white-light statistics for solar flares is limited, however.  As a result, intuition for such statistics from solar flares where the morphology of the active regions and the white light emission can be observed simultaneously is sorely needed \citep[][]{hudson1992}.

\section{Morphology of Complex Flares}
Beyond the aggregate measure of flare complexity shown in Figure \ref{fig:ncomp}, the detailed statistical morphology of complex stellar flares is largely unexplored, except for studies of individual large flare events \citep[e.g.][]{kowalski2010}. Deconstructing events into their constituent flare components offers a new opportunity to characterize the shape and composition of ``typical'' complex flares. This in turn may shed light on the origin of these complex events, such as from sympathetic flaring.

For example, if complex flares are typically the result of a single active region flaring repeatedly, we might expect that the largest energy component would occur first, or very near the beginning of the event. Each subsequent sub-flare in this ``cascade'' would then be smaller in energy, as the available magnetic energy in the active region decreased. However, if complex flares occur due to triggering of separate nearby active regions, then the sub-flares may have no preferred order in terms of energy. Instead, the temporal separation between the sub-flares could be related to the physical size scales between the active regions, as well as the stellar Alfv\'{e}n velocity.

\begin{figure}
\begin{minipage}[c]{0.6\textwidth}
\includegraphics[width=3in]{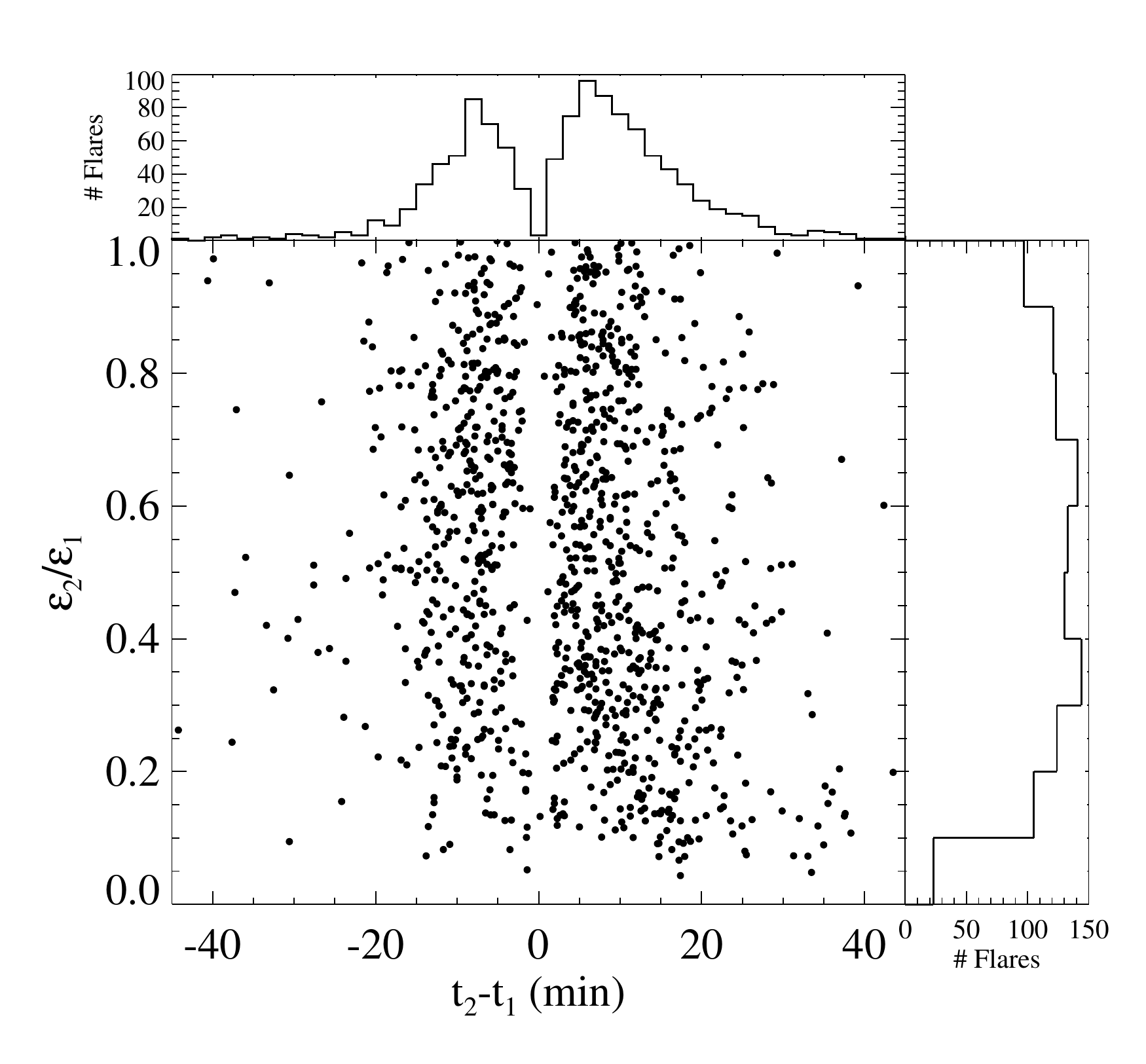}
\end{minipage}
\begin{minipage}[c]{0.35\textwidth}
\caption{Ratio of second-largest to first-largest component flare energies, versus the difference in their peak times. No clear trend is seen in the total component energy ratio distribution (right sub-panel). 61\% of second-largest components appear after the largest energy component within a flare event (top sub-panel), which corresponds to $t_2-t_1>0$.}
\label{fig:evt}
\end{minipage}
\end{figure}

The distribution of the energy ratios and time separations between the two largest sub-flares for each of the 1141 complex events from GJ 1243 is shown in Figure \ref{fig:evt}. For the majority of complex events, the largest sub-flare is observed before the second-largest, with 61\% of complex events having $t_2-t_1>0$. This is qualitatively consistent with the simple model of subsequent sub-flares depleting the magnetic energy in a single active region. However, given the slow decay profile of the flare template (Figure \ref{fig:model}), the $t_2-t_1$ distribution in Figure \ref{fig:evt} may partly be due to random superpositions of secondary sub-flares being more likely in the primary sub-flare's decay phase. A detailed Monte Carlo model of complex flares is needed to determine the significance of this result, and is beyond the scope of this work. The gap in flare separations near $t_2-t_1=0$ is due to a limited ability to resolve distinct sub-flares within a few minutes of each other, given the 1-minute sampling of the Kepler data. No preferred energy ratio ($\varepsilon_2/\varepsilon_1$) is found between the second-largest and largest sub-flares in these events, which qualitatively prefers the picture of complex flares occurring due to triggered flaring between multiple distinct active regions.

\section{Summary}
This sample of 6107 flare events from 11 months of Kepler short-cadence observations of the M4 dwarf GJ 1243 represents a unique benchmark dataset for studying flares from long duration light curves. This will be particularly valuable as a training dataset for flare detection algorithms (J. Davenport 2016, in preparation).
Direct comparison can also be made between this uniform sample of flares from a single star and ensemble flare samples from many stars \citep[e.g.][]{chang2015}.
To facilitate such future studies we have made the sample of flares, their measured properties, the complex flare determinations, and the corresponding 11 month light curve available online.\footnote{http://github.com/jradavenport/GJ1243-Flares}

The empirical white-light flare template has been used to robustly identify, and to some extent classify, complex flare events. In this work I have demonstrated a few simple examples of new ensemble statistics that are available for understanding complex white-light stellar flares. The physical origin of complex stellar flares may be revealed in the demographics and typical morphology of such events. Combining flare statistics with the magnetic topologies inferred from Zeeman Doppler Imaging \citep[e.g.][]{donati2008,morin2008b} may also help reconstruct a picture of the total magnetic activity on these stars. Analogous modern statistics for white-light solar flares, where the X-ray, UV, and optical emission can be traced with temporal-spatial detail, are needed to connect these stellar statistics with physical flare models.

\acknowledgements
JRAD is supported by an NSF Astronomy and Astrophysics Postdoctoral Fellowship under award AST-1501418.
Kepler was competitively selected as the tenth Discovery mission. Funding for this mission is provided by NASAÕs Science Mission Directorate.

\end{document}